# Unraveling Global Threads: Pandemic, Geopolitical Conflict, and Resilience in Fashion and Textile Supply Chain

**Md. AL-AMIN**
University of Massachusetts Lowell, USA;
alamintex20@yahoo.com

**Muneeb TAHIR**
North Carolina State University, USA
mtahir2@ncsu.edu

**Amit TALUKDER**
The University of Georgia, USA
at67309@uga.edu

**Abdullah AL MAMUN**
University of North Alabama, USA
amamun@una.edu

**Md Tanjim HOSSAIN**
orcid.org/0000-0002-2586-4775
North Carolina State University, USA
mhossai9@ncsu.edu ,

**Nigar SULTANA**
orcid.org/0000-0002-2586-4775
North Carolina State University, USA
nsultan3@ncsu.edu

**Abstract**

Several noteworthy scenarios emerged in the global textile and fashion supply chains during and after the COVID-19 pandemic. The destabilizing influences of a global pandemic and a geographically localized conflict are being acutely noticed in the worldwide fashion and textile supply chains. This work examines the impact of the COVID-19 pandemic, the Russo-Ukraine conflict, Israel-Palestine conflict, and Indo-Pak conflict on supply chains within the textile and fashion industry. This research employed a content analysis method to identify relevant articles and news from sources such as Google Scholar, the Summon database of North Carolina State University, and the scholarly news portal NexisUni. The selected papers, news articles, and reports provide a comprehensive overview of the fashion, textile, and apparel supply chain disruptions caused by the pandemic and the war in Ukraine, accompanied by discussions from common supply chain perspectives. Disruptions due to COVID-19 include international brands and retailers canceling orders, closures of stores and factories in developing countries, layoffs, and furloughs of workers in both retail stores and supplier factories, the increased prominence of online and e-commerce businesses, the growing importance of automation and digitalization in the fashion supply chain, considerations of sustainability, and the need for a resilient supply chain system to facilitate post-pandemic recovery. In the case of the Russo-Ukraine war, Israel-Palestine war, and Indo-Pak war, the second-order effects of the conflict have had a more significant impact on the textile supply chain than the direct military operations themselves. In addition to these topics, the study delves into the potential strategies for restoring and strengthening the fashion supply chain. It synthesizes insights from the latest literature to guide future research in understanding and addressing the challenges fashion supply chains face during this era. Overall, this study offers valuable insights that can inform future endeavors in researching and improving the fashion supply chain in the context of future pandemics and geopolitical conflicts.

### Küresel Bağlantıların Çözülüşü: Pandemi, Jeopolitik Çatışmalar ve Moda ile Tekstil Tedarik Zincirinde Dayanıklılık

### Özet

COVID-19 salgını sırasında ve sonrasında küresel tekstil ve moda tedarik zincirlerinde birkaç dikkate değer senaryo ortaya çıktı. Küresel bir salgının ve coğrafi olarak yerelleştirilmiş bir çatışmanın istikrarsızlaştırıcı etkileri, dünya çapındaki moda ve tekstil tedarik zincirlerinde keskin bir şekilde fark ediliyor. Bu çalışma, COVID-19 salgınının, Rusya-Ukrayna çatışmasının, İsrail-Filistin çatışmasının ve Hindistan-Pakistan çatışmasının tekstil ve moda endüstrisindeki tedarik zincirleri üzerindeki etkisini inceliyor. Bu araştırma, Google Akademik, Kuzey Carolina Eyalet Üniversitesi'nin Summon veritabanı ve akademik haber portalı NexisUni gibi kaynaklardan ilgili makaleleri ve haberleri belirlemek için bir içerik analizi yöntemi kullandı. Seçilen makaleler, haber makaleleri ve raporlar, salgının ve Ukrayna'daki savaşın neden olduğu moda, tekstil ve giyim tedarik zinciri kesintilerine ilişkin kapsamlı bir genel bakış sunuyor ve ortak tedarik zinciri perspektiflerinden tartışmalara eşlik ediyor. COVID-19 nedeniyle yaşanan aksaklıklar arasında uluslararası markaların ve perakendecilerin siparişleri iptal etmesi, gelişmekte olan ülkelerdeki mağaza ve fabrikaların kapanması, hem perakende mağazalarında hem de tedarikçi fabrikalarda çalışanların işten çıkarılması ve ücretsiz izne çıkarılması, çevrimiçi ve e-ticaret işletmelerinin artan önemi, moda tedarik zincirinde otomasyon ve dijitalleşmenin giderek artan önemi, sürdürülebilirlik hususları ve pandemi sonrası toparlanmayı kolaylaştırmak için dayanıklı bir tedarik zinciri sistemine duyulan ihtiyaç yer alıyor. Rus-Ukrayna savaşı, İsrail-Filistin savaşı ve Hindistan-Pakistan savaşı durumunda, çatışmanın ikinci dereceden etkileri, doğrudan askeri operasyonların kendisinden daha önemli bir tekstil tedarik zinciri etkisine sahip olmuştur. Bu konulara ek olarak, çalışma moda tedarik zincirini onarmak ve güçlendirmek için potansiyel stratejileri araştırmaktadır. Bu dönemde moda tedarik zincirlerinin karşılaştığı zorlukları anlamak ve ele almak için gelecekteki araştırmalara rehberlik etmek üzere en son literatürden içgörüler sentezlemektedir. Genel olarak, bu çalışma gelecekteki pandemiler ve jeopolitik çatışmalar bağlamında moda tedarik zincirini araştırma ve iyileştirme konusundaki gelecekteki çabaları bilgilendirebilecek değerli içgörüler sunmaktadır.

**Anahtar Kelimeler**: COVID-19; Rusya-Ukrayna savaşı; Tedarik zinciri; Moda; Tekstil; Pandemi; İsrail-Filistin savaşı; Hindistan-Pakistan savaşı; Giyim; E-ticaret.

### Introduction

The fashion industry, known for its dynamic nature and intricate global supply networks has been severely impacted by the COVID-19 pandemic. The ensuing disruptions have spurred a significant reorientation of research efforts within academia and industry alike. Traditionally, research in the fashion supply chain encompasses a broad range of topics including sustainability, consumer behavior, production processes, and global sourcing strategies. However, the COVID-19 and Russia-Ukraine conflict has necessitated a fundamental shift in focus toward critical health-related concerns, global supply crisis, and supply chain resilience.

According to Plunkett (2020), research trends in the global textile and apparel supply chain in 2020 mainly focused on the effect of COVID-19 on the global fashion industry, reshoring, moving fashion orders from China to low-wage countries, smart textiles, additive manufacturing, robotics, green fashion, automation, and fashion rental services (Plunkett, 2020). According to Plunkett and Plunkett Research (2019), one of many research trends right before the pandemic revolved around the evolution of the supply chain. Before 2019, the sustainable fashion supply chain and fast fashion were emerging as new trends (Cetinguc et al., 2017). However, during COVID-19, research trends turned to qualitative research and non-empirical works on the impact of COVID-19 on the fashion supply chain (Chowdhury et al., 2021). Despite the rampant supply chain disruptions amid the pandemic, the textile industry saw large-scale production activity in the personal protective clothing segment. However, implementing lockdowns and industry closures worldwide resulted in significant strain on the global textile and apparel supply chain (WTO, 2021). This strain was further compounded by the ongoing worldwide shipping crisis and frequent mutations in the SARS-CoV-2 virus, necessitating government responses through repeated lockdowns. The recent Russian invasion of Ukraine has added another layer of complexity to the situation, exacerbating the challenges faced by the textile supply chain. Although the Russo-Ukrainian war has had a relatively minor *direct* impact on the textile supply chain, with reduced supply and demand from the combatant states, the *second-order effects* have caused actual disruptions. These adverse *second-order effects* include trade embargoes, higher energy and prices, global inflation, US federal reserve's quantitative tightening and a stronger dollar, monetary tightening, and dwindling global demand.

Geopolitical conflicts significantly disrupt the textile and fashion supply chain by hindering trade routes and increasing shipping costs and times. These conflicts also lead to volatility in energy and raw material prices, alongside reduced consumer demand, exposing the supply chain to increased vulnerability (Bednarski et al., 2025; Rasshyvalov et al., 2024). The impact of the Russo-Ukrainian war, Israel-Palestine conflict, and Indo-Pak war on the global textile supply chain cannot be explained without detailing the context in which it evolved in a highly interconnected world (Figure 1). This conflict emerged on the horizon when the world finally recovered from the Covid-19 pandemic-induced demand slump. The reverberations of weakened global demand are rippling through the textile supply chain, resulting in multifaceted implications encompassing logistics, cost, and the availability of textiles.

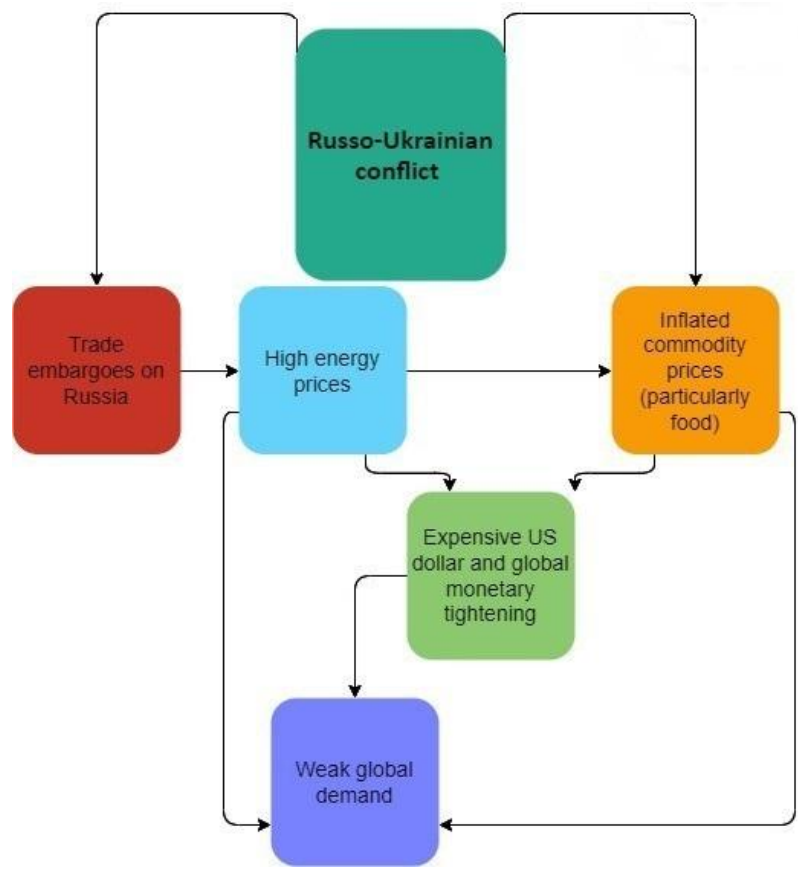

Figure 1. A visual representation of how a chain reaction of second-order effects has weakened global demand, including textiles.

(Bednarski et al., 2025; Rasshyvalov et al., 2024). Trend analysis is pivotal in informing future research and corporate business plans by shaping the fashion industry and identifying the most viable supply chain models. On the other hand, content analysis provides valuable contemporary insights and critical information on specific areas or issues (Biondi et al., 2020; Ibrahim et al., 2015). This study investigates the impacts of the COVID-19 pandemic and other notable geopolitical conflicts across the world including the Russo-Ukraine war, Isarael-Palestine conflict, and Indo-Pak war on the overall textile and fashion business worldwide through a systematic content analysis and regular literature review based on the textile and fashion supply chain.

## 1. Literature Review

A supply chain is a complex system that includes raw materials, sourcing, transportation, production, distribution, and delivery to the end users. The global supply chain is even more complex than the domestic supply chain as it also includes worldwide sourcing and distribution that are affected by several factors, such as international trade rules, tariffs, taxes, and location (Ibrahim et al., 2015). Supply chain disruption occurs due to unexpected natural or human events created in the upstream network, inbound logistic network, or the sourcing environment that disrupts the normal activities of the firm (Golan et al., 2020). There are several notable supply chain disruptions in the recent past, such as the 2011 earthquake in Japan, measuring nine on the Richter scale resulted in a destructive tsunami and subsequent accident at the Fukushima nuclear power plant. This natural disaster had far-reaching impacts on various industries, including disrupting car parts production in Japan. As a direct consequence, the General Motors truck plant located in Louisiana, USA, experienced a halt in production due to the unavailability of Japanese car parts (Golan et al., 2020).

Moreover, the epidemic caused by the Ebola virus in a few African countries in 2014 disrupted the overall food supply chain resulting in an acute food scarcity across that region. For the same reason, the raw rubber industry in Liberia dropped its production significantly, creating a crisis in rubber-dependent industries across Africa (Al-Amin et al., 2024; Sumo, 2019). Throughout the last century until today, the world has faced several devastating pandemics caused by deadly viruses. These include the infamous "Spanish Flu" (1918-1919), the "Asian Flu" (1957-1958), the "Hong Kong Flu" (1968), the outbreak of SARS-CoV-1 (2002-2003), and the emergence of the "Swine Flu" (2009-2010). These pandemics have resulted in an immense loss of lives worldwide and have severely disrupted domestic, regional, and global supply chains (Koch et al., 2020a). Similarly, geopolitical conflicts in the 20th century, such as the Arab states' oil embargo after the Arab-Israel War of 1973 and the temporary slump in oil supplies during the Gulf War, significantly disrupted fossil fuel-dependent industries in developed and developing nations respectively (Lieber, 1992; *Milestones*, 2022).

The textile and apparel supply chains constitute a complex and dynamic network. From its inception to the final delivery to end-users, finished apparel undergoes numerous stages within this intricate network. These stages include fiber production, spinning, knitting/weaving, design, cutting, sewing, dyeing/printing, incorporation of accessories, packaging, testing, shipping, warehousing, retailing, and many others. Each stage plays a vital role, and any disruption within this chain significantly threatens the entire system's functionality and efficiency (Hui & Choi, 2016). Even a likelihood of disorders creates greater vulnerability in this complex supply chain network (Golan et al., 2020). The COVID-19 outbreak had significant and varied effects on the worldwide supply chain, affecting individual and collective economic activities in unprecedented ways (Guan et al., 2020). During March 2020, the pandemic impacted approximately 80%-95% of the global supply chain (Remko, 2020). This impact was seen in consumer behavior, production, logistics, services, retailing, and social and environmental sustainability, shifting the supply chain ecosystem to a transformative paradigm (Mollenkopf et al., 2021). Textile and apparel manufacturing are labor-intensive industries, and developing countries are the predominant suppliers of fashion goods in Europe and North America (Castañeda-Navarrete et al., 2021). The sudden outbreaks of COVID-19 jeopardized the manufacturing activities in developing countries resulting in an unprecedented FSC disruption worldwide (Kabir et al., 2021). In Europe during April-June 2020, clothing production fell by 37.4% compared to April-June 2019, and the clothing retail business experienced a sales drop of 43.5% (Pasquali, 2021).

Alternatively, when considering the impact of the Russo-Ukrainian war on the global textile and apparel supply chain, one should recognize that despite not being preeminent textile and apparel industry players, Russia and Ukraine contribute to the global apparel market. Russia's high demand for clothing makes it the world's tenth-largest importer of apparel products. At the same time, Ukraine's proximity to the European market allows it to cater to a portion of the manufacturing demand of some of the top Western brands (*Textile – UkraineInvest*, 2023; WTO, 2021). There are indications that the punitive sanctions imposed on Russia following its invasion of Ukraine have a noticeable impact on its economy. Specifically, there appears to be a reduction in domestic consumption and a decline in imports (Sonnenfeld et al., 2022). These developments are also retarding apparel exports to the world's tenth-largest apparel importer (Caglayan, 2022; Mirdha, 2023). In India, few apparel exporters reported disruptions in timely payments from Russia due to tanking of the Russian Ruble against the US Dollar in the early days of the Russo-Ukraine war (Sinha, 2022). Meanwhile, Vietnamese exporters have reported difficulties ensuring deliveries to Russian customers in the early months of the Russo-Ukraine war (Industry of Vietnam, 2022).

While the downturn in Russian apparel imports has had a relatively minor impact on the world's major apparel exporters, as the Russian market constitutes a smaller portion of their overall export basket, it has left countries like Uzbekistan, which heavily rely on textile exports to Russia, in a precarious economic situation (GET, 2022; IMF, 2022). Russia seems to have found a viable conduit to access the global market through China. By leveraging its strong economic ties with China, Russia has provided its importers with continued access to merchandise, including textiles, and conducted trade in yuan, the Chinese currency. This strategic move has not only helped Russia bypass the impact of the sanctions but also presented a unique opportunity for other countries to explore alternative channels to enter the Russian market. For instance, Bangladesh has expressed keen interest in tapping into the Russian market by dealing in yuan, opening up new avenues for trade and commerce (Investment Monitor, 2022; Mirdha, 2023).

Now, it is a critical moment to evaluate the overall impact of the pandemic and the geopolitical upheaval in Eurasia on various textile and apparel supply chain paths, aiming to enhance resilience and strengthen these supply chains against future crises.

2. **Research Objectives**

A scarcity of studies investigating the combined impact of the COVID-19 pandemic and the Russo-Ukraine war on the global fashion, textile, and apparel supply chains at a macro level has been identified. Very few studies have focused on the combined effect of these two subsequent crises. Most of those discussed only on global supply chain crises rather than solely focusing on textile and fashion supply chain (Allam et al., 2022; Cekerevac & Bogavac, n.d.; Minh Ngoc et al., 2022; Saleheen et al., 2024). Only one paper talked about the disruptions of textile supply chain due to this dual setback (Chowdhury, 2024). Therefore, this study aims to examine and analyze the available body of peer-reviewed research and news about the distortions in the fashion, textile, and apparel supply chains during the peak period of the COVID-19 pandemic and the Russo-Ukraine war. This research aims to contribute to the existing scholarly discourse on this topic and enhance the understanding of the complexities and challenges these supply chains face during these successive crises.

### 3. Methods

This study follows a content analysis approach involving three key steps: content identification, explanation of identified content categories, and identification of research gaps (Ibrahim et al., 2015). The content analysis methods employed were peer-reviewed research analysis, news article analysis, governmental agency reports analysis, international governance institutional reports analysis, and international financial institutional reports analysis. The author thoroughly analyzed research articles to identify key themes. The identified vital themes were then refined to select the most significant ones. Finally, these top themes were further examined using the latest literature and news sources relevant to each theme.

Google Scholar was utilized with the search terms "Fashion Supply Chain ANDCOVID-19" to comprehensively search scholarly articles from January 2020 to March 2021. Google Scholar and North Carolina State University Network's "Summons" portal were used with the phrase "Textile and Apparel Supply Chains AND Russia-Ukraine War." Google Scholar provides a broader search scope by examining full-text content from scholarly sources, distinguishing it from databases like PubMed and Web of Science that only focus on abstracts. Given the pandemic's recent emergence, a wide net was cast to capture as many relevant studies as possible. The initial search yielded 45 scholarly articles, subsequently filtered based on the relevance of the Fashion Supply Chain (FSC) to COVID-19, resulting in 17 pertinent peer-reviewed research articles. We analyzed these articles for themes, identifying common themes

A search for the news articles, governmental reports, international governance institutional reports, and international financial institutional reports was performed through "Nexis Uni" scholarly news content and the "Summons" portal through North Carolina State University network using the terms "fashion supply chain AND COVID-19" for 2020-2021, specifying the location of the United States and "Textile and Apparel Supply Chains AND Russia-Ukraine War."  The search result brought 92 miscellaneous articles from the abovementioned sources, which were further filtered and narrowed down to 40. These 40 articles were thoroughly analyzed for themes related to FSC, its impact on textile and apparel supply chains in the aftermath of the Russo-Ukraine war, and COVID-19. Finally, the top FSC themes are listed in

### 4.  Themes found in content analysis

The results  indicated that the textile, apparel, and fashion retail and manufacturing industries came to a standstill during the pandemic and continue to face significant headwinds following the initiation of armed conflict in Ukraine. The fashion retail sector witnessed substantial sales declines, leading major brands and retailers to cancel numerous orders, impacting production in developing countries and resulting in layoffs, job losses, and factory closures. The implementation of lockdown measures and restrictions on public gatherings compelled individuals to stay at home, leading to a surge in online clothing orders and a subsequent boost to the e-commerce sector. Studies investigating supply chain resilience during global health crises, such as the COVID-19 pandemic, emphasized the importance of automation and digitalization. The Russo-Ukraine war has caused fluctuations in global energy markets, while inflationary pressures in several economies have prompted central banks worldwide to adopt demand compression measures. These measures have contributed to subdued demand for textile and apparel products.

**Table 1**. Themes found in content analysis (research articles).

| SL | Keywords | References |
|---|---|---|
| 01 | Decline in sales | (Brydges et al., 2021; Chakraborty & Biswas, 2020; Kabir et al., 2021) |
| 02 | Order cancellation | (Brydges et al., 2021; Carta Mantiglia Pasini, 2020; Castañeda-Navarrete et al., 2021; Chakraborty & Biswas, 2020; Hibberd & Ali, 2021; Kiilunen & Ferrara, 2020; McMaster et al., 2020) |
| 03 | Job loss, layoff, factory closure | (Brydges et al., 2021; Castañeda-Navarrete et al., 2021; Hibberd & Ali, 2021; Kiilunen & Ferrara, 2020) |

| 04 | Online shopping, e-commerce | (Casini & Roccetti, 2020; Kvrgic et al., 2020; M & Kannappan, 2020; Petković et al., 2020) |
|---|---|---|
| 05 | Automation, digitalization | (Casini & Roccetti, 2020; Kiilunen & Ferrara, 2020) |
| 06 | Supply chain resilience, Domestic consumption and imports contraction, Freight rates increase (sea and rail transport), food supply disruptions, Soaring inflation, crude undersupply, Geopolitical Risks, inflationary pressures, declining GDP growth | (Casini & Roccetti, 2020; Kiilunen & Ferrara, 2020; Kvrgic et al., 2020; McMaster et al., 2020) |

**Table 2**. Themes found in content analysis (miscellaneous articles)

| SL | Keywords | References |
|----|----------|------------|
| 01 | Sustainability | (Ell, 2020; GEF, 2021; Stower, 2020b) |
| 02 | Order cancellation, wage, and payment cut | (Roberts-Islam, 2021; Stower, 2020a; Uddin, 2020) |
| 03 | Job loss, layoff, factory closure, store closure | (Kurtenbach, 2020; Roberts-Islam, 2021; Stower, 2020a; Uddin, 2020) |
| 04 | Online shopping, e-commerce | (GoSourcing LLC, 2020; Stower, 2020a) |
| 05 | Automation, digitalization | (Burstein, 2021; Kathiala, 2020; Product News Network, 2020; Stower, 2020b) |
| 06 | Supply chain resilience, vulnerability, transparency, supply bottlenecks, production problems, inflation | (Burstein, 2021; Ell, 2020; Kurtenbach, 2020; Stower, 2020b) |

**Results**

İndicate that the global fashion supply chain increasingly prioritizes environmental sustainability, with fashion brands embracing green and circular fashion practices to mitigate ecological impacts.

Moreover, COVID-19 disrupted textile sustainability by increasing demand for disposable personal protective equipment (PPE) and synthetic fibers that led to higher plastic waste and reduced recycling. Lockdowns halted many sustainable initiatives, slowed supply chains, and shifted consumer focus from eco-friendly choices to affordability and hygiene. However, the crisis also raised awareness about resilient and circular systems by encouraging innovation in sustainable materials and local sourcing (Annaldewar et al., 2021; Haukkala et al., 2023; Yasmeen et al., 2022).The COVID-19 pandemic restricted public gatherings and movement, reducing retail store sales. During strict lockdowns, clothing stores remained closed, significantly decreasing work orders for suppliers in developing countries. Many fashion brands responded by canceling existing orders, leading to job losses and layoffs for workers in apparel factories. Following the Russo-Ukraine war, reports of textile and apparel inventory buildups have emerged due to order cancellations or postponements. Manufacturing costs have risen due to import control regimes, increased energy, and raw material input costs.

### 4.1. Analyses of the themes found

The discussion section is subdivided into two sections. The first section discusses the themes found while the literature review centered around the COVID-19 pandemic's impact on the fashion supply chain, while the second section covers the themes present in textile and apparel supply chain effects due to the Russo-Ukraine war.

### 4.1.1  Effects of the COVID-19 on the Fashion Supply Chain

The effects of the recent pandemics are listed below according to the literature review and most notable themes found.

#### 4.1.1.1. Store Closure and Sales Decline

The COVID-19 pandemic inflicted significant costs on international textiles, fashion, and apparel brands and retailers across the globe. Major companies faced immense difficulties maintaining their normal operations, primarily due to stay-at-home orders that resulted in temporary store closures and substantial revenue declines. To understand the specific impact on brands and retailers in the United States, WWD investigated the repercussions of COVID-19 on this sector (Ilchi, 2020). A few are listed (Table **3** ).

**Table 3.** Financial impact due to COVID-19 on the major USA fashion brands and retailers

| SL | Company | Period | Financial Impact |
|---|---|---|---|
| 1 | Abercrombie & Fitch Co. | | 34% sales drop, $244.2 million net loss |
| 2 | American Eagle Outfitters | Three months ending May 2, 2020 | 45% sales drop, $334 million revenue fall. |
| 3 | H&M | | 57% sales drop, March 1 to May 7, 2020 |
| 4 | J.C. Penney Co. | Up to May 15, 2020 | Filed bankruptcy, 152 stores closed, 1000 employee layoff |
| 5 | Kohl's Corp | Three months ending May 2, 2020 | 40.60% sales drop, $541 million loss |
| 6 | Levi Strauss & Co. | Three months ending May 2, 2020 | 15% workforce cut, 62% revenue fall. |
| 7 | Macy's | For the quarter ending May 2, 2020 | 45% sales decline, 3900 jobs cut |
| 8 | Nike Inc. | March to May 2020 | $790 million sales loss |
| 9 | PVH Corp. | For the first quarter, 2020 | Net loss of $1.1 billion |
| 10 | Ralph Lauren | For the quarter ending June 27, 2020 | Revenues fall 65.9% |

**Source:** Ilchi, 2020.

At the beginning of the pandemic, Boston Consulting Group (BCG) forecasted an estimated sales drop for luxury brands, which would be between $85 to $120 billion in 2020. Fashion and all other luxury categories were staring at a sales decline of $450-$600 billion. The report also reported a sales drop of 85% in China during the first two-month lockdown, which was 95% in some European countries such as Italy, Spain, and France (Biondi et al., 2020). Retailers were compelled to comply with government orders and close their doors to consumers during the COVID-19 pandemic. Consequently, consumer spending experienced a continuous decline, which took time to recover. Online apparel brands, including

Amazon, witnessed a significant sales decrease of 42% on average between mid-February and mid-March 2020 (Baum et al., 2020). Other major retail brands, such as Abercrombie & Fitch, American Eagle Outfitters, H&M, J.C. Penney, Macy's, and Ralph Lauren, suffered terribly from the demand destruction unleashed by the COVID-19 pandemic (see Table **3**). Abercrombie & Fitch recorded a 34% sales drop and incurred a net loss of $244.2 million over three months. J.C. Penney filed for bankruptcy, leading to the closure of 125 stores and the layoff of 1000 employees. Levi Strauss and Ralph Lauren experienced revenue declines of 62% and 65.9%, respectively. American Eagle Outfitters and Macy's witnessed sales declines of 45% and 57%, respectively. According to the Center for Retail Research (CRR), 2020 was the worst year for retail businesses in the past 25 years, resulting in a significant loss of High Street jobs. The pandemic led to nearly 180,000 retail job losses in 2020 (Haigh, 2021). However, Forbes predicted 2021 would witness more fashion business bankruptcies, although there is a high chance of recovery. Notably, middle-market brands were expected to disappear. In contrast, luxury and low-market brands were slated to survive in the long run.

### 4.1.1.2. Order Cancellation and Joblessness

After the rapid store closures in May and April 2020, most large brands canceled orders or delayed payment to their overseas suppliers (Reuters, 2020). According to Business Insider, from April to June 2020, USA brands and retailers cut the orders to half the same time as the previous year, equivalent to approximately $9.70 billion. In contrast, European brands and retailers reduced orders by $6.50 billion. Together, they refused to pay overseas suppliers of more than $16 billion orders since the beginning of the pandemic. This situation dealt a severe blow to suppliers in developing countries, resulting in the closure of numerous businesses and the unfortunate layoffs of millions of factory workers (McNamara, 2020). Several developing countries, including Cambodia, Albania, Myanmar, Indonesia, Pakistan, Sri Lanka, and certain countries in Central America, encountered significant challenges in the aftermath of the pandemic. The consequences are manifested as permanent or temporary factory closures due to widespread order cancellations. Notably, Bangladesh emerged as one of the most affected countries, witnessing a distressing impact on its garment industry. The country experienced a high number of factory closures, leading to an inability to pay wages to approximately 1.14 million workers (Chakraborty & Biswas, 2020). In May 2020 alone, Bangladesh experienced a substantial decline in orders from US brands and retailers, resulting in a loss of $355 million in order value, representing a 45% decrease compared to regular order placements (Dean, 2020). Consequently, nearly one million workers in Bangladesh were either dismissed or furloughed, exacerbating the economic hardship faced by these individuals. Similarly, in Cambodia, over 400 factories suspended their production as of July 2020, leading to approximately 150,000 workers losing their jobs (ILO, 2020). Many jobless workers had insufficient food reserves at home, leading to acute food insecurity during this period. The situation raised concerns about the potential for a silent famine to emerge, especially if the government imposed further lockdown measures (Kabir et al., 2021). However, following intense controversy, criticism, protests, and media coverage, fashion brands eventually reversed their decisions partially or fully.

**Table 4.** Order status of major fashion brands with Asian apparel suppliers

| SL | Fashion Brands | Order Status |
|----|----------------|--------------|
| 01 | C&A | Canceling all orders until June 2020, later promised to pay 93% of finished and running orders and negotiated the remaining 7%. |
| 02 | Next Plc | Canceled some orders; will pay the finished and running orders up to April 2020. |
| 03 | ASOS Plc | Canceled orders of $6-$7.3 million initially; promised to pay for all orders. |
| 04 | Mango | Will pay for only finished goods; will delay the payment for running orders. |
| 05 | New Look | Canceled 20% of orders for summer and spring from Bangladesh. |
| 06 | J.C. Penney | Delays payments for already placed orders. |
| 07 | H&M | Will pay for all orders and place new orders. |
| 08 | Inditex | Will pay for all orders. |
| 09 | M&S | Will pay for all the orders that shipped before March 2020. |
| 10 | Tesco Plc | Committed to paying for all orders |

**Source:** Reuters, 2020

If the pandemic persists or newer one comes, it is anticipated that several adverse consequences will befall developing countries, where apparel manufacturing serves as a primary source of income for the lower-income class, including permanent job losses, increased reliance on short-term contracts, wage reductions, and unpredictable incomes. Such circumstances would render necessities such as housing, food, and other essential supplies unaffordable for affected workers, exacerbating socio-economic and humanitarian crises (ILO, 2020). Additionally, the financial crisis deepening alongside the struggle for survival will compel the global fashion industry to prioritize the welfare of workers directly linked to profit margins. The order status of prominent brands and retailers from the USA and Europe during the initial phase of the pandemic is presented in Table 4. Notably, brands such as C&A, Next, ASOS, and New Look initially canceled either a significant portion or the entirety of orders placed with their suppliers. However, many of these brands eventually reversed their order cancellations following frequent criticisms from stakeholders and human rights organizations. Nevertheless, this critical period of the pandemic had credibility costs for the buyers regarding ethical responsibilities and financial obligations to suppliers. These developments underscore the complex dynamics and ethical challenges faced by the fashion industry, particularly during times of crisis. It highlights the urgent need for responsible and sustainable practices to safeguard the well-being and rights of workers throughout the fashion supply chain (Freeman et al., 2022).

### 4.1.1.3. Online and E-commerce Business

The impact of COVID-19 extends beyond the fashion manufacturing industry, as it has significantly influenced the buying behaviors of retail consumers and various aspects of retail businesses. Prolonged lockdown measures confined individuals to their homes, prompting them to seek alternative methods of purchasing essential goods, leading to a remarkable transformation in global buying trends. Online shopping has experienced a notable surge, driven further by the unprecedented COVID-19 pandemic (Bhatti et al., 2022). Even small retailers who had previously not ventured into online sales quickly adopted temporary online sales options through social media platforms, implementing services like curbside pickup or home delivery (Koch et al., 2020b). However, despite the growing popularity of online shopping, it still carries certain risks associated with the delivery process. Parcels passing through various hands and channels during postal delivery make them susceptible to becoming potential contagion carriers (Silvestri, 2020). A prominent online data provider, Statista, reported that in May 2020, 62% of adults in the US expressed disinterest in shopping at physical stores, while 52% showed a greater inclination toward online shopping. Additionally, 20% of individuals significantly increased their frequency of online purchases, and even those who had never previously considered online shopping began embracing this method (Pasquali, 2021). Consequently, this shift in consumer buying behavior has resulted in uncertain demand patterns and supply chain challenges within the e-commerce industry.

It is worth noting that this change in buying patterns has occurred globally, affecting regions such as Asia, Europe, Africa, and the Americas. As a result, countries that were not prominent players in the pre-pandemic situation have emerged as significant participants in the e-commerce landscape (Abdelrhim & Elsayed, 2020).

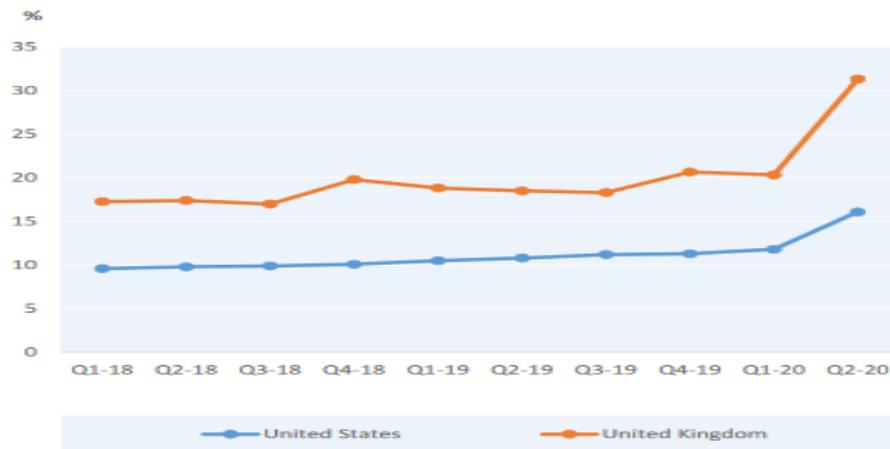

**Figure 2**. Ecommerce market share in total retail sales in the USA and UK from 2018-2020. Source:OECD, 2020

According to OECD the e-commerce market share in the total retail business witnessed a gradual increase in the United States between the first quarters of 2018 and 2019, growing from 9.6% to 11.8%. However, this growth experienced a significant surge, reaching 16.1% in the first and second quarters of 2020 (OECD, 2020). Similarly, the e-commerce market share in the United Kingdom followed a similar pattern. It rose from 17.3% in the first quarter of 2018 and 20.3% in 2020. However, during the first and second quarters of 2020, there was a substantial spike, reaching 31.3%. The e-commerce market share in China was 24.6% in August 2018, followed by a decrease to 17.3% in January 2020. However, it rebounded and reached 24.6% again by August 2020.

### 4.1.1.4. Automation and Digitalization

The COVID-19 pandemic has laid bare vulnerabilities in the global fashion supply chain, highlighting the importance of digitalization and automation for better resilience during crises. Businesses already implementing digital and automated systems demonstrated greater adaptability during this challenging period. As a result, the significance of automation and digitalization has gained prominence in efforts to revive the fashion industry and supply chain in a post-pandemic scenario (Gonzalo et al., 2020). The pandemic has accelerated the pace of fashion supply chain transformations that were initially projected to occur five to six years later. Embracing automation and digitalization in the supply chain offers numerous advantages over traditional methods, including cost efficiency, speed, transparency, accuracy, and customization (Just Style, 2020). Small businesses must adopt new operating models and technologies to navigate the post-pandemic recovery. This adaptation process will require enhancing their capabilities, infrastructure, and resilience (Dua et al., 2020).

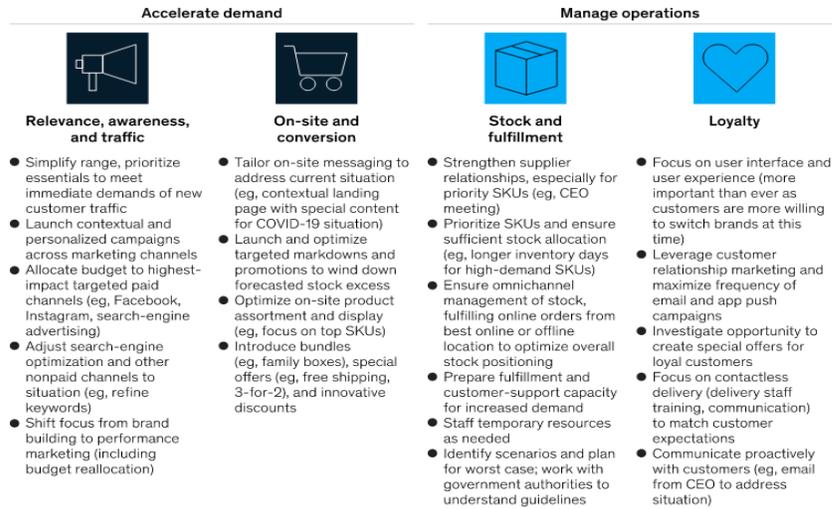

**Figure 3**. Options for digital  transformation in fashion retail( Gonzalo et al., 2020)

Silvestri (2020) discussed the conveniences that automation and digitalization could provide to the global fashion supply chain in assisting the supply chain during the long-lasting pandemic (Silvestri, 2020).

- Virtual Reality (VR) and Augmented Reality (AR) technologies offer valuable solutions for addressing size and fit concerns when shopping for apparel online, consequently reducing return rates. Once online consumers become accustomed to this technology, they will likely feel more confident in their purchases and may even be encouraged to buy more. A notable example is GAP's AR app, "DressingRoom," which utilizes Google Tango to allow customers to conveniently test the size and fit of clothing items from any location and at any time.

- Artificial Intelligence (AI) can be harnessed to identify data patterns and trends, enabling accurate sales forecasting. AI also facilitates the monitoring of user opinions, automates tasks, and helps create personalized customer experiences. By leveraging AI technology, the entire fashion value chain can be modernized and disrupted, revolutionizing the post-pandemic fashion retail business, especially in e-commerce. AI can increase production efficiency in the textile and apparel industries by reducing reliance on human labor by implementing robotics. This automation can accelerate reshoring and reindustrialization efforts in developed countries, where renowned brands and retailers are often based (Al-Amin et al. 2023). However, one criticism of this thrust of automation is job losses in the labor-intensive industries in the developing world (Emont & Journal, 2018). Others have argued that automation does not cut down apparel industry jobs but improves access to job opportunities in businesses employing automation by amplifying their productivity (Parschau & Hauge, 2020). Additionally, AI can be utilized for product quality inspection, ensuring higher standards and minimizing errors. Furthermore, it has the potential to enhance working conditions, making workplaces safer and more conducive for employees. The combination of AI and blockchain technology promises complete transparency throughout the global fashion supply chain in the post-pandemic fashion business. By leveraging blockchain's decentralized and immutable nature alongside AI's data analysis capabilities, brands, and consumers can gain complete visibility into product origins, supply chain processes, and sustainability practices.

### 4.1.1.5. Supply Chain Resilience and Transparency

Resilience indicates risk management and recovery in an effective way from a disrupted situation. A resilient supply chain can restore itself to its previous state or even surpass it (Hsu et al., 2021). Supply chain transparency involves easy access to comprehensive product information, encompassing the entire journey from raw materials to the finished product (Hänninen, 2020). The fashion supply chain has been severely impacted by the pandemic, necessitating appropriate actions to adapt to the changed post-pandemic landscape. Researchers have extensively studied supply chain risks, emphasizing risk identification, assessment, mitigation, and monitoring as crucial components of supply chain risk management. However, strategies such as agility, robustness, flexibility, redundant capacity, and surplus inventory alone are insufficient to ensure resilience in the post-pandemic fashion supply chain (Fahim et al., 2020). Hosseini et al. (2019) proposed a concept of resilience capacity in the supply chain, comprising three capacities (Hosseini et al., 2019).

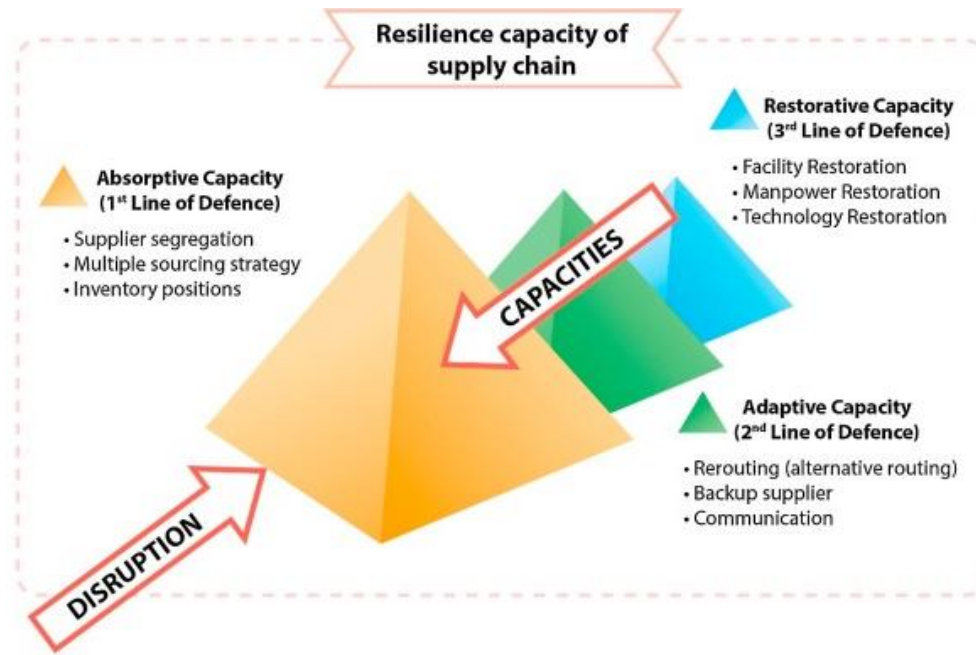

**Figure 4.** Resilience Capacity of supply chain (Hosseini et al., 2019)

1. Absorptive capacity – Absorptive capacity is the first line of defense that minimizes a system's ability to absorb or endure the impact of system disruptions and minimalizes the adverse effects of that disruption with a minor expenditure of energy or effort.

2. Adaptive capacity – Adaptive capacity is the second line of defense which indicates the degree to which a system can cope with and try to overcome the disruption by employing modified operating practices deprived of any recovery activities.

3. Restorative capacity – Restorative capacity is the third line of defense. It indicates the ability of a system to be reestablished rapidly and effectively when the absorptive and adaptive capacities of that system cannot maintain the accepted level of performance.

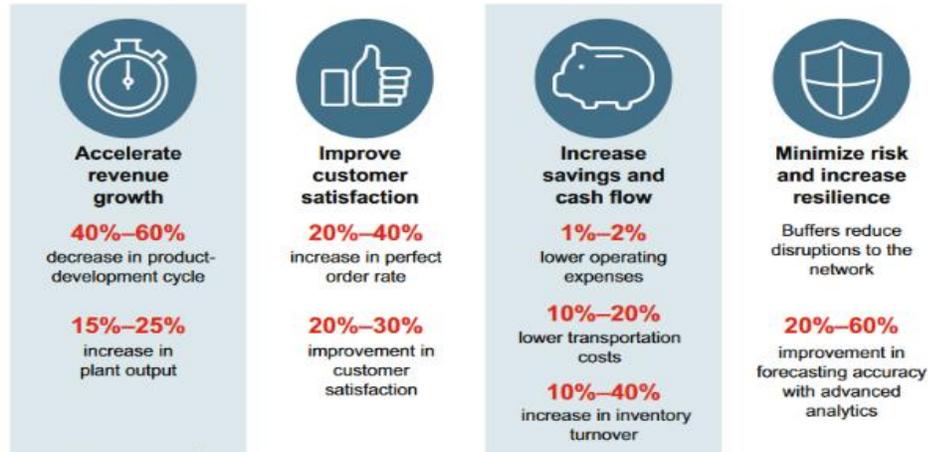

**Figure 5.** Benefits of supply chain resilience (Schatteman et al., 2020)

Schatteman et al. (2020) discussed the Bain analysis highlighting the advantages of a resilient supply chain. According to the analysis, resilient supply chains experience faster growth by effectively meeting market demands and satisfying consumers' needs (Schatteman et al., 2020). These resilient supply chains can increase their order rate by 20%-40% and improve customer satisfaction by up to 30%. Additionally, they achieve cost savings through enhanced cash flow and inventory turnover, with a potential increase of 10%-40%. Companies that invest in building supply chain resilience can also reduce product development time by 40%-60% and optimize operations to maximize production capacity by 15%-25%. In a study by Sharma et al. (2021) focusing on retail supply chain business determinants for post-pandemic resilience, collaboration efficiency emerged as the top determinant for accelerating performance (Sharma et al., 2021). Other determinants identified include order fulfillment, inventory control, digital retail supply chains, diversification and offshore strategies, collaborative planning, maintaining demand, capacity, and supply, and shifting to e-commerce.

### 4.1.1.6. Impact of Geopolitical Conflicts

There are several conflicts have been taking place across the different regions around the world after the severe aftermath of COVID-19, such as Russia-Ukraine war, Israil-Palestine conflict, and recently India-Pakistan war. As a result, textiles and fashion supply chain have been disrupted worldwide significantly. All those events have been studied and discussed in this section based on the available primary and secondary sources.

### 4.1.2. Impact of Russo-Ukraine War on Textile and Apparel Supply Chain

Russia-Ukraine war has been putting great impact on world's economy over the last several years. It has been reviewed from different literature to find out the relevant keywords. Those are listed in Table 5 and discussed the findings afterwards.

**Table 5.** Impact of Russo-Ukraine War on Textile and Apparel Supply
Chain

| Keywords | References |
| --- | --- |
| Supply bottlenecks; Production problems; Inflation increase | (Strauß, 2023) |
| Increase in gas and fuel energy prices | (Horton & Palumbo, 2022) |
| Supply bottlenecks | (McKenna, 2022) |
| Freight rates increase (sea and rail transport); Food supply disruptions | (Ngoc et al., 2022) |
| Domestic consumption and imports reduction | (Sonnenfeld et al., 2022) |
| RMG exports drop; financial institutions, e.g., SWIFT ban | (Mirdha, 2023) |
| Production order suspension; Soaring energy prices | (Caglayan, 2022) |
| Piling Stocks; Hold shipments | (The Times of India, 2022) |
| Increase prices of raw materials; Limit capital mobilization | (Industry of Vietnam, 2022) |
| Export reduction; Remittance drop | (GET, 2022) |
| Trade disruptions, Higher fuel prices | (IMF, 2022) |
| Financial sanctions | (Investment Monitor, 2022) |
| Tight material supply | (VietnamPlus, 2022) |
| Reduction of natural gas production | (Agnolucci et al., 2023) |
| Interrupted supply of gas and power | (Islam, 2022) |
| Soaring inflation; Fuel shortage; | (Varadhan et al., 2023) |

| Reduction of exports of cotton yarn; | (Apparel Resources, 2023) |
|---|---|
| Declining production revenue | |

### 4.1.2.1. Curtailment of financing to the Russian war machine

After Crimea's annexation in 2014, the United States and its allies imposed economic sanctions on Russia. In response to the Russian invasion of Ukraine in 2022, the United States and its partners implemented new, more wide-ranging sanctions targeting the Russian economy. Using economic sanctions to isolate Russia from its traditional export markets and financial services was intended to achieve strategic goals. The overarching strategic goals were twofold: to significantly reduce the resources supporting Russia's military campaign in Ukraine and to mobilize domestic public opinion against the conflict by inducing economic hardships through the sanctions. While the efficacy of these sanctions in achieving their strategic objectives remains hotly debated, the impact of the sanctions on global supply chains is a widely recognized phenomenon that has created significant disruptions in the international trade landscape.

### 4.1.2.2. Commodity supply shock and high energy prices

Military conflict, being a disastrous event by nature, is exacerbated in the current scenario as it involves two nations that are among the world's most prolific commodity exporters. The untimely occurrence of this armed conflict is deplorable as it coincided with a global commodity super cycle, which had already burdened populations worldwide with inflationary pressure caused by deficiencies in supply chains at the height of the Covid-19 pandemic crisis.

The forces of supply and demand govern the fundamental workings of the global economy. The global economy embarked upon a path to recovery once COVID-19 pandemic-related restrictions began relaxing around the globe with increasing immunization rates. As global economic growth gained momentum, so did the demand for fossil fuels to power that growth, escalating the cost of crude oil globally. Russo-Ukrainian war amplified the already high energy costs owing to Russia's position as one of the foremost non-OPEC energy exporters in the world and efforts to cut this revenue spinner for the Russian state to size.

Various wet processes involved in textile manufacturing, such as sizing, de-sizing, dying, and finishing, are highly energy-intensive and directly reliant on fossil fuels such as natural gas or electricity generated from fossil fuels (Palamutcu, 2015). Upswings in global energy prices make textile processing an expensive proposition. On the textile raw materials level, the dependence of textiles on fossil fuel-derived energy pricing trends could not be understated. Synthetic fibers like Polyester, Nylon, Acrylic, and Polypropylene are derived from crude oil, while approximately two-thirds of world fiber production comprises synthetic fibers (A. A. Mamun et al., 2022; Nagalakshmaiah et al., 2019; Suaria et al., 2020). The escalation of global crude oil prices has resulted in a spill-over effect on the costs of polymers used as raw materials for textile fiber production. Increasing fuel commodity prices affect synthetic textile fibers and natural fibers such as cotton. Cotton is the most widely cultivated textile fiber in the world. Its agricultural production and processing heavily depend on fossil fuels (Derby et al., 2022; Harri et al., 2009), utilized for mechanized farming equipment, fertilizer manufacturing, and crop transportation from farms to ginning facilities..

Following the Russian invasion of Ukraine, the world experienced a severe natural gas crisis, prompting European nations to shift away from Russian gas consumption. In response, cash-rich European economies turned to limited liquefied natural gas (LNG) supplies, causing prices to skyrocket in the LNG spot purchase market. As a result, net energy-importing economies such as Pakistan and Bangladesh could not compete in the market for LNG spot purchases, leading to power outages in these countries and adversely affecting their traditional textile and apparel production (Agnolucci et al., 2023; Dilawar, 2022; Islam, 2022; Varadhan et al., 2023). Cotton yarn producers in Gujarat, India, who trade with Eurozone textile manufacturers, are experiencing reduced product demand due to higher energy costs, leading to factory closures (Apparel Resources, 2023; Talukder et al., 2022).

With natural gas prices soaring and a limited supply in the market, nitrogen-based fertilizers manufactured from natural gas and essential for managing crop yields of plant fiber and edibles experienced a significant price spike (Baffes & Koh, 2022). The global fertilizer supply chain faced additional shocks when Russia, the world's most prolific exporter of fertilizers (Statistica, 2023), restricted exports of this crucial agricultural input in response to sanctions complicating a problematic situation further (Baffes & Koh, 2022).

Truncation of Russian supplies in the energy markets has led to exorbitant energy price rates. A documented direct relationship exists between rising energy prices and inflation in the broader global economy (Venditti, 2013). The spike in textile and apparel manufacturing inputs and food inflation increased due to rallying energy prices.

### 4.1.2.3. Response to rising global inflation

The confluence of food supply shock and elevated energy prices has caused alarm in financial capitals worldwide. Central banks have begun to combat inflationary tendencies in highly interdependent world economies. The US federal reserve's quantitative tightening to moderate demand pressures in the US economy has highlighted the necessity of taking decisive action, and many central banks worldwide have followed suit. Acts of the US federal reserve have ushered in an era of a stronger US dollar while wooing international investments in high return-yielding US treasury papers. Investment into developing economies has seen a downturn while drying up another source of foreign exchange inflows required to maintain foreign exchange buffers. As net food importers, textile and apparel exporters like Pakistan, Sri Lanka, Egypt, and Bangladesh have been battered by the Russo-Ukraine war (Lau, 2022; A. Mamun et al., 2022; Nature, 2022; Rahman & Uddin, 2023). The disruptions in the supply chains and the stronger US dollar have led to higher costs of financing food, energy, and other imports, eroding these countries' foreign exchange buffers. As a result, they have turned to the International Monetary Fund (IMF) to rebuild their foreign exchange reserves. However, IMF programs are often accompanied by austerity measures, which may involve withdrawing special privileges from specific industries, including textiles and apparel, making them less competitive than their international counterparts.

Textile and apparel exporting economies of Turkey, Bangladesh, Pakistan, and Egypt have experienced a devaluation of their currencies, making raw material imports expensive, increasing the prices of manufactured textiles and apparel products, and further eroding their competitive advantage. Pakistan implemented import controls to mitigate the rapid depletion of its foreign exchange reserves caused by costly imports and inadequate dollar inflows. These controls have had a detrimental impact on the textile industry. The industry's plans for modernizing and importing machine spares and raw materials are currently stalled at the country's ports (Dilawar, 2023; Inam, 2023). Due to the rising cost of living courtesy of rampant inflation in these countries, the labor costs in these economies' textile and apparel sectors are bound to rise to make them less competitive. Increasing input and operational costs require businesses to either factor in these realities in the pricing regimes of their merchandise or sacrifice a portion of their profits to retain market share and consumer interest. The former strategy is fraught with the risk of dampening consumer demand, while the latter's persistence leaves little room for further expansion and capacity building.

#### 4.1.2.4.   Demand compression

US federal reserve projects a 1.5% loss in global GDP and a 1.3 percentage point rise in inflation due to the Russo-Ukraine war (Caldara et al., 2022). There is a weak demand for goods and services in an environment of monetary tightening worldwide to manage inflationary expectations. Inflation also suppresses the demand for apparel as consumers become more circumspect about spending (Hamdan et al., 2022). Major apparel and textile-producing countries like India, Vietnam, and Bangladesh have reported declining exports due to dampening global demand. In Bangladesh and Cambodia, apparel exports are not growing as briskly as before the Russo-Ukraine war (Fashionating World, 2023; Reed, 2022). Meanwhile, textile and apparel exports in India and Vietnam have fallen due to the *second-order effects* of the Russo-Ukraine war (Kumar & Jadhav, 2022; TEXtalks, 2022).

#### 4.1.3.   Impact of Israel-Palestine conflict

The Israel-Palestine conflict is one of the longest and most intractable territorial-political disputes of the modern era that has significantly influenced the political landscape of the Middle East while challenging crucial aspects of the global economy, trade flows, and investor sentiment worldwide. Rooted in competing nationalist claims dating back to the early twentieth century, this conflict has created ongoing socio-economic challenges including disruptions to global supply chains (Khare et al., 2025). These geopolitical tensions not only impact regional security but also affect the flow of international trade through the Suez Canal. Whenever conflict upheavals occur, the potential for disruption to shipping traffic increases and creates uncertainty for global trade players (Aisyah Hamzah et al., 2025). The Suez Canal is a vital waterway that connects the Red Sea and the Mediterranean Sea. It handles about 12% of the world's shipping, 30% of all container trade and 13.5% of the world's freight. Various research indicates that maritime transport is essential for global supply chain connections and economic growth. Any disruption to Suez Canal can have significant consequences and costs for various stakeholders in the supply chain. For example, the blockage of the canal in March 2021 during the COVID-19 pandemic, highlights the vulnerability of global supply chains (Man-Yin Lee et al., 2021; Wan et al., 2023). Disruptions to trade routes or increased instability in the region hinders the movement of raw materials, production, and finished goods. This leads to delays, increased costs, and potential disruptions in the supply chain for textile companies that rely on the Middle East, specifically Jordan and Egypt (Çitil et al., 2025). Additionally, this conflict has led to the disruption of the transit of raw materials needed for the textile and fashion industry both for middle east and other world. The rerouting of shipping vessels has isolated Aqaba, Jordan's only seaport from key international textile trade routes (Grüneisl & Labadi, 2024). Fashion and textiles buyers of Europe and America are trying to avoid this chaotic route resulting in moving their orders from countries, such as Bangladesh, India, and Pakistan to Turkiye (Mirdha, 2024). The renowned fashion blog shenglu.com pointed out the following impacts of Red Sea attacks due to Israil-Palestine conflict on the global fashion and textile supply chain (Sheng, 2024).

- Global shipping rates have surged with routes from Asia to Northern Europe more than doubling since December 2023. Retailers in the U.S. and EU face steep cost hikes (for example, 700–700–1,800 per container) forcing brands to either absorb expenses or pass them to consumers.

- Shipments to the U.S. East Coast and Europe now take 12–14 extra days due to rerouting via the Cape of Good Hope. EU retailers cite 3–4 weeks delay, while Chinese and Indian exporters face 7–10 and 12–14 days setback disrupting inventory planning and increasing lead times for seasonal collections.

- Chinese exporters are pivoting to rail freight (despite price hikes) to avoid sea delays. EU brands resist airfreight due to carbon concerns exacerbating reliance on slower, congested routes. These alternatives further strain budgets and operational flexibility.

- Suppliers in Bangladesh, Pakistan, and India are absorbing higher freight and insurance costs eroding margins. EU brands fear bearing the brunt of delays while Pakistani exporters face losses by honoring pre-conflict pricing agreements amid soaring charges.

Delays in textile machinery and raw material imports (e.g., for Pakistan's textile sector) threaten factory slowdowns. Bangladesh's dependency on Red Sea routes for 70% of exports risks production halts compounding challenges for just-in-time manufacturing models

### 4.1.4. Impacts of India-Pakistan Conflict

As mentioned earlier, India and Pakistan are the major producers of textile raw materials, such as cotton. Moreover, they are also the major exporters of textiles and fashion to EU and North American countries. The India-Pakistan conflict is a complex and enduring issue, primarily centered on territorial disputes, historical grievances, and religious nationalism (Hussain & Ali Naqvi, 2025). Recent escalations following the Pahalgam attack and then retaliatory missile actions by India have led a full range war between these two nations (CNN, 2025). This has already brought the noticeable impact on their regional maritime commerce, forcing shipping companies to divert shipments bound for Karachi Port to other locations. Exporters and logistics providers are reporting higher costs and extended delivery times as the conflict disrupts the main regional shipping routes. For example, major Chinese logistics companies COSCO and OOCL have suspended their operations to the region. Additionally, OOCL also introduces a substantial rate increase of $500 per container for shipments still routed to Karachi (Shipping News, 2025). Bangladesh buy their textile raw materials, such as cotton, yarn, and dyestuffs mostly from these two countries is now eyeing for alternative sources as anticipating the duration of this conflict to be a longer like Russo-Ukraine conflict (Report, 2025). Textile and fashion brands that rely on Indian or Pakistani suppliers for raw materials or completed products are currently dealing with prompt shipping delays and escalating expenses, particularly affecting fashion labels, such as Zara and H&M that

source both cotton materials and finished clothing from this region (Maclang, 2025).

## 5.     Future Recommendations

Based on the above review, the following future recommendations are being proposed by the authors.

- Adopting strategies that enhance absorptive, adaptive, and restorative capacities to cope with any future disruptions. Focusing not only on inventory and flexibility but also on long-term risk management systems.

- Utilizing Artificial Intelligence (AI), automation, and blockchain to improve supply chain efficiency, transparency, and traceability. Technologies, such as big data and predictive analytics are recommended to modernize operations and reduce returns.

- Collaborating with suppliers and stakeholders for better forecasting, capacity planning, and crisis response to incorporate multi-tiered contingency plans for future disruptions.

- Expanding online sales platforms and omni-channel logistics to remain resilient against demand shocks and store closures.

- Building systems that allow for full traceability from raw materials to end products to reduce reliance on geopolitically unstable regions. Encouraging regional production, nearshoring, and reshoring to increase supply chain robustness.

## 6.      Limitations

Despite its valuable insights, this study has limitations that should be acknowledged. One notable limitation is the reliance on Google Scholar as the primary source for scholarly articles. Although efforts were made to narrow the search to scholarly content, using a machine algorithm for article identification rather than human curation introduces the possibility of missing relevant articles. Furthermore, due to the relatively minor role of Russia and Ukraine in the global fashion, textile, and apparel manufacturing sectors, there is a lack of peer-reviewed studies explicitly evaluating the impact of the conflict between these two countries in the broader supply chain. To compensate for this dearth of literature, this study had to supplement scholarly research with information from governmental sources, multilateral donor agencies, international governance institutions, and news media reports to construct a comprehensive understanding of the effects of the conflict on the fashion supply chain. It is essential to acknowledge the spatiotemporal constraints of this work. As the COVID-19 pandemic and the war in Ukraine unfold, more research articles and news reports are expected to emerge in scientific peer-reviewed journals and news portals. Moreover, the Israel-Palestine and India-Pakistan conflicts on textile supply chain have very limited number of sources. Incorporating these additional sources in future editions of this study would enhance the robustness of the findings and provide a more comprehensive analysis.

## 7.    Conclusion

This study explored the themes linked to fashion, textile, and apparel supply chain disarray due to the COVID-19 pandemic and the Russo-Ukraine war in the research literature and news media. The findings of this study could serve as the foundation for future content analyses, which may uncover additional themes as time progresses. The findings for COVID-19 showed that order cancellation could be a big factor during pandemic for fashion garments and textiles producers. Due to the financial crisis, both apparel retailers and producers had to make factory closures and layoffs which resulted in humanitarian crises in the developing countries. E-commerce played a big role in satisfying consumers' needs during this pandemic as online was the only option to order for the needed textiles. As there were restrictions on public gatherings, it was difficult to find the required workforce to continue the production. This finding raises the demand for digitalization and automation in the textile and fashion industry. The findings due to pandemic ultimately showed the fragility of the then supply chain demanding more resilience and adaptability. The overall findings highlight the urgent need for industry to develop resilient strategies such as diversified sourcing, investment in agile technologies, and stronger supplier relationships. The sector must prioritize adaptability, transparency, and ethical labor practices to mitigate risk in case similar crisis arises in future.

The findings on the impact of Russo-Ukraine conflict on textile and fashion supply chain pointed out the crisis of energy, such as crude oil and natural gas to produce textiles, apparel, and garments in the EU countries. Europe's dependency on Russian energy led to a surge in electricity and gas prices that drastically raises production and transportation costs for textile manufacturers. Economic sanctions against Russia limited trade movements, especially affecting companies reliant on Russian suppliers or markets. Major fashion brands ceased operations in Russia and disrupted retail and e-commerce markets in the region that impacting global revenue streams. With instability and inflation, consumer spendings in the affected regions declined resulting in the reduced demand for fashion and textile products. The conflict heightened awareness of geopolitical vulnerabilities and encouraged companies to diversify sourcing and rethink global supply chain strategies. Moreover, we have presented the latest scholarly viewpoints on how the fashion supply chain can restore its shape and strengthen in the aftermath of the COVID-19 pandemic. While Russia and Ukraine may not have a significant role in the global fashion supply chain, the conflict between these nations has contributed to energy market volatility and posed a threat to post-pandemic recovery. The conflict between these two countries has made energy markets more volatile and threatened the post-pandemic recovery. Inflationary pressures are building up in the world economy, and demand compression measures in the global north and import suppression impulses in the global south have adversely affected the demand and supply dynamics of fashion, textile, and apparel products.

On the other hand, the Israel-Palestine and Indo-Pak conflicts provided a crucial extent of vulnerability against textile and fashion supply chain. The middle eastern conflict often compels the shipping companies to reroute their transit that leads to increase delivery time and cost of products. Also, middle eastern textile economy has been impacted badly as Jordan faces closure of their main seaport. On the other hand, the Indo-Pak conflict causes EU and USA retailers to rethink their decision of sourcing from South Asia due to frequent geopolitical conflicts. The fashion and textile sector must proactively build more adaptable and transparent supply chains to navigate future global disruptions and safeguard vulnerable workers. This requires a continued focus on diversified sourcing, technological adoption, and reinforced ethical practices to ensure long-term resilience and sustainability in the face of unforeseen crises.